\documentclass[aps,10pt,twocolumn]{revtex4}
\usepackage{amsfonts}
\usepackage{amsmath}
\usepackage{amssymb}
\usepackage{graphicx}

\begin{document}
\newcommand{\ve}[1]{\mbox{\boldmath $#1$}}
\newcommand \beq{\begin{eqnarray}}
\newcommand \eeq{\end{eqnarray}}
\newcommand \bea{\begin{eqnarray}}
\newcommand \eea{\end{eqnarray}}
\newcommand \beo{\begin{equation}}
\newcommand \eeo{\end{equation}}
\def\simge{\mathrel{%
      \rlap{\raise 0.511ex \hbox{$>$}}{\lower 0.511ex \hbox{$\sim$}}}}
\def\simle{\mathrel{
      \rlap{\raise 0.511ex \hbox{$<$}}{\lower 0.511ex \hbox{$\sim$}}}}

\title{The Rayleigh-Taylor Instability in a Bose-Einstein Condensate}
\author{Swarbhanu Chatterjee}
\affiliation{Department of Physics, University of Rhode Island,
Kingston, RI 02881 - 0817} 

\begin{abstract}
We show how a two dimensional Bose-Einstein Condensate trapped in
a nonequilibrium state can be driven into the Rayleigh-Taylor
instability if an outward in-plane force is exerted on it. If the
condensate is inside a semiconductor, the above force could arise
from an inhomogeneous strain applied on the host semiconductor
crystal. On the other hand, for a BEC of alkali atoms, the force
could be exerted by an inhomogeneous magnetic field turned on
after the BEC has been released from its magnetic trap. During its
expansion, the condensate will break into droplets each a separate
BEC. Therefore, one can create BEC droplets out of a single large
BEC and study the coherence properties of the droplets with
respect to each other. The discussion on the linear onset of the
Rayleigh-Taylor instability in a BEC is generalized to three
dimensions.
\end{abstract}


\vskip0.0pc]

\startpage{1}
\maketitle

\section{Introduction}

The Rayleigh-Taylor instability ( RT ) \cite{rayleigh,chandra} is
observed in fingers that develop on industrial smoke spreading
from factory chimneys \cite{baym1}, blast remnants of supernovae
\cite{ribeyre}, superfluid droplets from charged superfluid jets
\cite{chun}, and in exciton droplets in bulk Ge \cite{baym}. It
occurs in inhomogeneous fluids with density gradients when a force
acts in the direction of less density. Layers of constant density
become unstable towards the formation of undulations or wiggles.
Since a wiggle displaces mass in the direction of the force, a
constant density layer which has undulations has a lower potential
energy than one which is straight \cite{baym,sharp}. The
wavelength of the wiggles is bounded from below by the surface
tension or the viscosity, since there is a greater cost in surface
energy for a wiggle of shorter wavelength and therefore, greater
surface area, while viscosity does not permit adjacent fluid
elements to have very different velocities as is needed for short
wavelength undulations. A necessary condition for the RT
instability to evolve in a fluid is that it should take a shorter
time for sound to propagate through the medium than the time in
which the instability grows. Therefore, $s \gtrsim \lambda/\tau$
must hold, where $s$ is the sound speed in the fluid, while
$\lambda$ and $\tau$ are the wavelength and growth time of the
instability.

Ever since BECs were first created in the laboratory, experiments
in BECs, particularly on their collapse and revival
\cite{greiner1}, the quantum interference of two BECs
\cite{andrews} and the superfluid to Mott-Insulator transition in
optical lattices \cite{greiner2} have justified the hope that the
laws of quantum mechanics can be studied in such systems. BECs
made out of atoms are trapped in magnetic traps and are physically
isolated from the walls of the vacuum chamber. This limits
decoherence processes to interactions with uncondensed particles
which remain in the vicinity of the condensate. On the other hand,
BECs in semiconductor systems usually have a shorter decoherence
time, $\tau_d$, compared to their atomic counterparts because of
greater interaction with the host, since generally $\tau_d \sim
\tau_e$, where $\tau_e$ is the interaction time \cite{zurek}. BECs
in semiconductors trapped in more than one dimension are truly
extended systems with the host environment making quantum
mechanical measurements of the density of the condensate
wavefunction at different parts of the condensate, hence quickly
destroying the long-range order. The onset of decoherence through
the destruction of long range order as above is related to the
Kibble-Zurek mechanism \cite{kibblezurek}, which is important in
cosmology. The presence of additional channels of decoherence
within a solid due to a greater interaction with the host makes it
more difficult to make BECs in solids that can survive long enough
to be observed. Magnon condensation has been reported in the
magnetic compounds, Cs$_2$ Cu Cl$_4$ and Tl Cu Cl$_3$ \cite{radu}.
A solid state system, which is an interesting but not yet proven
as a candidate system for Bose-Einstein Condensation is a cold
excitonic cloud trapped in a coupled quantum well (CQW)
\cite{butov2,snoke,snoke2}.

When it comes to BECs, hydrodynamic phenomena play a very
important role. For example, vortices in rotating BECs are
essentially hydrodynamic in nature. To make safe inferences about
the quantum mechanics, it is necessary to have an understanding of
hydrodynamic instabilities, such as the RT instability, that may
develop in them in certain experimental conditions. Furthermore,
hydrodynamic instabilities may give rise to new situations in
which quantum mechanics can be further tested. In this paper, we
show how in certain conditions a Rayleigh-Taylor instability may
develop in a BEC giving rise to the formation of droplets. These
droplets, each of which is a BEC, can have interesting properties
that need future study. One interesting possibility is to make
clones of a parent BEC and then make quantum mechanical operations
on each of them separately.

The Rayleigh-Taylor instability in BECs has not yet been
investigated. This paper aims to address that issue. In Sec. II,
we consider a two dimensional BEC that can be unstable towards RT
instability. Later, we shall see that our results can easily be
generalized to three dimensions. In Sec. III, we describe the
linear onset of the RT instability. In Sec. IV, we remark on the
future evolution of the instability. In Sec. V, we discuss two
possible BEC systems in which we might see the formation of
Rayleigh-Taylor droplets.

\section{The Rayleigh-Taylor Instability in a BEC}

\subsection{ Initial State of the 2D BEC at T=0}

We consider a two dimensional BEC of excitons at rest in a
harmonic trap of frequency, $w$. The initial potential is $V(r)
\simeq w^2r^2/2$. The Thomas-Fermi (TF) density of the exciton
cloud, $\rho_0 \sim [A-r^2]^{1/2}$ at $r< \sqrt (\hbar/m w)$, with
$A = ( 2 \mu / m w^2)$ and the chemical potential, $\mu$.

We now consider the release of the trap and in its place, the
immediate application of a radial force outward from the former
trap's center. We call this as time, $t=0$. The condensate is in
non-equilibrium at $t=0$. We may neglect the curvature of the
expanding BEC as long as the wavelength of the instability is much
smaller than the radius at which it sets in. To describe our
system, we choose the y axis as being perpendicular to the surface
of the condensate, with the surface at $y=0$. We assume that the
cloud expands in the $\hat y$ direction, and that the instability
develops in the $x$-plane in the neighborhood of $y=0$.  We assume
that the density of the condensate is linear at the surface where
the instability happens. In other words,
$\rho_0=\rho_1=\textrm{constant}$ at $y<-\delta$, $\rho_0 =
\rho_1(1 - y / \delta)$ for $-\delta < y < \delta$, and
$\rho_0=\rho_2=$ constant at $y> \delta$. We are studying the
case, $\rho_1
> \rho_2$. We also assume that the density varies slowly over
time, $\partial \ln \rho_0/
\partial ~t \ll 1/\tau$, where $\tau$ is the time of growth of the
instability. This is true if $s>>\lambda/\tau$, in which case the
instability evolves slower than the time it takes sound to
propagate \cite{chatterjee}. The above condition involving the
speed of evolution of the instability is necessary for an RT
instability to develop in any system.

\subsection{Onset of the Rayleigh-Taylor instability in a BEC at
T=0}

The hydrodynamic equation of motion for a Bose-Einstein condensate
at T=0 is, \beo \frac{\partial \vec{v}}{\partial t} = - \frac{\nabla
P}{m\rho} - \frac{\hbar^2}{2m^2} \nabla \Big\{ \frac{\nabla^2
\sqrt{\rho} }{\sqrt{\rho}} \Big\} - \frac{\nabla U}{m} -\nabla
\Big(\frac{\vec{v}^2}{2}\Big),
 \eeo
 where U, the external potential, includes both trap and applied potentials,
$P=\rho \mu$ is the pressure, with $\mu$, the chemical potential.
When all the forces are perfectly balanced, the condensate is in
static equilibrium, \beo -\nabla
P_0-\frac{\hbar^2\rho_0}{2m}\nabla \Big\{\frac{\nabla^2
\sqrt{\rho_0}}{\sqrt{\rho_0}} \Big\} + m \rho_0 \vec{g}=0, \eeo
where the subscript $0$ denotes unperturbed quantities and
$m\vec{g}=-\nabla U$ is the total external force on each particle.
Then, at $t=0$, the trap is released and immediately a radial
outward force is applied. Therefore, from $t=0$ onwards, the
condensate is in non-equilibrium. The condensate starts expanding
due to the new potential gradient. We shall not consider the
excitations in the system, but will only study the hydrodynamic
flow of the part that remains a condensate. We note that in a
force field, the lowest energy state is not the zero momentum
state.

 We now consider the linear stability of
the condensate against perturbations in which the fluid is locally
displaced infinitesimally by $\delta \vec{r}\,(x,y,t)$. We
describe the disturbance by local changes in the density,
$\rho(x,y,t) = \rho_0(y,t)+\delta\rho(x,y,t)$, the pressure,
$P(x,y,t)= P_0(y,t)+\delta P(x,y,t)$, and the velocity, $\vec
u(x,y,t) = \delta \vec u\,(x,y,t)= \partial \delta
\vec{r}\,(x,y,t)/\partial t$. The BEC behaves locally as an
incompressible fluid, and we may assume that the density $\rho$ of
a given volume element of the condensate remains constant as the
element is displaced by $\delta \vec r\,(t)$.
\begin{eqnarray}
 \rho(\vec{r},t+\delta t) &=& \rho_0(\vec{r}-\delta \vec{r}\,(t), t)
 \\ \nonumber
 &=& \rho_0(\vec{r},t)- \nabla \rho_0\cdot \delta \vec{r}\,(t),
\end{eqnarray}
so that the perturbation of the density is given by
\begin{equation}
\frac{\partial \, \delta \rho}{\partial t} = -\delta \vec {u} \cdot
  \nabla \rho_0.
\label{a1}
\end{equation}
The continuity equation, \beq \frac{\partial \rho}{\partial t} +
\nabla \cdot (\rho \vec u\,) = 0. \label{continuity0} \eeq when
linearized becomes, \beq \frac{\partial \, \delta \rho}{\partial
t} + \nabla \cdot (\rho_0\delta \vec{u}) = 0; \label{continuity}
\eeq leading to
\beq \nabla \cdot \delta \vec{u} =0. \label{a2}
\eeq The linearized equation of motion for the condensate with
only the lowest order terms in the perturbation is,
\begin{equation}
m\rho_0 \frac{\partial \delta \vec{u}}{\partial t} = -\nabla
\delta P+ m\delta \rho\vec{g} -\frac{\hbar^2}{4m}\nabla(\nabla^2
\delta\rho). \label{leom}
\end{equation}
The line of discontinuity in density, $\rho_0$, will become
slightly deformed in the perturbed state. Following
\cite{chandra}, let the line be then defined by $y_s+\delta y_s$,
with \beo  \delta y_s (x,t)=\delta y_s(0) e^{i(kx-wt)}\eeo. The
discontinuity in normal stress required by equilibrium is $T_s
d^2\delta y_s/dx^2$, where $T_s$ is the `effective tension' of the
condensate, which we define as the kinetic energy per length along
the locus of the density drop, similar to the surface tension
defined in Ref. \cite{khawaja}. \beo T_s =
\frac{\hbar^2}{2m}\int_{-\delta/2}^{\delta/2} dy |\nabla \psi|^2
\simeq \frac{\hbar^2}{4 m \delta }
\frac{(\rho_1-\rho_2)^2}{(\rho_1+\rho_2)}, \label{tension} \eeo
assuming a linear fall in the density at the defect over a surface
width, $\delta$. Therefore, we should add a term corresponding to
the effective tension to the right side of Eq.\,(\ref{leom}),
which leads to, \beq m\rho_0 \frac{\partial \delta
\vec{u}}{\partial t} = &-&\nabla \delta P+ m g \delta \rho\hat{y}
-\frac{\hbar^2}{4m} \nabla(\nabla^2 \delta \rho)
\nonumber \\
&-& \hat{y} k^2 \tau T_s \delta u_y  \delta(y), \label{eee} \eeq
where we have used $d\delta y_s/dt = \delta u_y(y=0)$, and
therefore, $\delta y_s= \tau \delta u_y(y=0)$.

Following Chandrashekhar \cite{chandra}, we look for an instability of transverse
wavelength $k$
such that any perturbed quantity, $Q$, grows in time as,
\begin{equation}
 \delta Q(x,y,t) = Q'(y) e^{ikx + t/\tau}.
\end{equation}
Taking the x-component of Eq.\, (\ref{eee}), \beo \frac{ m \rho_0
\delta u_x}{\tau} = -ik \delta P- i\frac{\hbar^2 k}{4 m} \nabla^2
\delta \rho. \label{ox} \eeo Since, from Eq.\, (\ref{a1}), \beo
\delta \rho = - \tau \frac{d\rho_0}{dy} \delta u_y,
\label{deltarho} \eeo is nonzero only at $y=0$, solving
Eq.\,(\ref{ox}) for $y \neq 0$ gives \beo
 \frac{m \rho_0 \delta u_x}{\tau} = -ik \delta P.
\label{leox}
\eeo
Similarly the y-component of Eq.\,(\ref{eee}), for $y \neq 0$, is,
\beo
\frac{m \rho_0 \delta u_y}{\tau}=-\frac{d \delta P}{dy}.
\label{leoy}
\eeo
Differentiating Eq.\,(\ref{leox}) with respect to x leads to
\beo
\frac{m \rho_0}{\tau} \frac{d \delta u_x}{dx}=k^2 \delta P.
\label{ux}
\eeo
Eqs.\,(\ref{a2}) and (\ref{ux}) lead to,
\beq
-\frac{m \rho_0}{\tau} \frac{d \delta u_y}{dy}=k^2  \delta P.
\eeq
Differentiating with respect to y, and using Eq.\,(\ref{leoy}), we find,
\beo
-\frac{m \rho_0}{\tau} \frac{d^2 \delta u_y}{dy^2}=k^2 \frac{d \delta P}{dy}
= -\frac{m \rho_0 k^2}{\tau} \delta u_y.
\eeo
Therefore,
\beo
\Big( \frac{d^2 }{dy^2} -k^2 \Big) \delta u_y= 0.
\label{inst}
\eeo
The condensate is unstable to velocity perturbations of the form
$\delta u_y=A e^{t/\tau}e^{\pm ky+ikx}$.
With boundary conditions $\delta u_y,d
\delta
u_y/dy \to 0$ as $y \to \pm\infty$, the general solution of Eq.~(\ref{inst})
with an interface at $y=0$ is,
\begin{eqnarray}
 \delta u_y &=& A e^{ky}, \quad y<0
\nonumber \\
 &=& A e^{-ky}, \quad y>0
\label{uy}
\end{eqnarray}
with $k>0$, since $\delta u_y$
 must be continuous across the interface. It should be noted that
 $\delta u_x$ need not be
since the condensate is inviscid.

To determine the solution and thus the growth of the instability, we
require the boundary conditions across the interface.
Taking the y-component of Eq.\,(\ref{eee}), we find,
\beq
\frac{d \delta P}{dy} &=& - \frac{ m \rho_0 \delta u_y}{\tau} -
\tau m g \delta u_y \frac{d\rho_0}{dy} - k^2 \tau T_s \delta u_y \delta(y)
\nonumber \\
& & ~ + \frac{\tau \hbar^2}{4 m}\frac{d}{dy} \Big\{ \nabla^2 \Big(
\frac{d \rho_0}{dy} \delta u_y \Big) \Big\}. \label{B} \eeq
Integrating Eq.\,(\ref{B}) from $y=0-$ to $0+$ gives, \beq \Delta
\left[\delta P \right]&=& - \Delta \left[ m g \tau \delta u_y \rho
\right]
 - k^2 \tau T_s \delta u_y(y=0)
\nonumber \\
& & ~ + \frac{\hbar^2 \tau}{4 m} \Delta \left[ \nabla^2
\Big(\frac{d \rho_0}{dy}u_y \Big) \right], \label{deltaB} \eeq
where $\Delta$ denotes the difference across the interface. On the
other hand, Eqs.\, (\ref{a2}), (\ref{ox}) and (\ref{deltarho})
lead to, \beo \Delta \left[\delta P \right] = - \frac{m }{k^2
\tau} \Delta \left[ \frac{d u_y}{dy} \rho_0 \right],
\label{deltaA} \eeo since $d\rho_0/dy=0$ for $ y \neq 0$, and
therefore, $\Delta[\nabla^2(u_yd\rho_0/dy)]=0$. Eliminating
$\Delta(\delta P)$ from Eqs.\, (\ref{deltaB}) and (\ref{deltaA}),
we find the boundary condition, \beo
 \Delta\left[\frac{m }{k^2 \tau}  \frac{d u_y}{dy} \rho_0\right]=
\Delta \left[ m g \tau \delta u_y(y=0) \rho_0 \right]+ k^2 \tau
T_s \delta u_y(y=0), \eeo which, after substitution for $\delta
\vec{u}_y$ using Eq.\, (\ref{uy}), and Eq.\, (\ref{tension}),
gives the relation between the growth time of the instability and
the wavevector, \beo
\tau=\frac{1}{k^{1/2}}\left[g\Big(\frac{\rho_1-\rho_2}{\rho_1+\rho_2}\Big)
-\frac{\hbar^2 k^2}{4m^2
\delta}\Big(\frac{\rho_1-\rho_2}{\rho_1+\rho_2}\Big)^2
\right]^{-1/2}. \label{tau} \eeo For $\rho_1>\rho_2$, the solution
is unstable ($\tau$ is real and positive) for $0<k<k_c$ where $
k_c=\Big\{ m (\rho_1 - \rho_2 )g/T_s\Big\}^{1/2} $. Using
$d\tau/dk=0$ in Eq.\, (\ref{tau}) gives us the fastest growing
mode, which has a wavevector
 and growth time,
\beo
\tilde{k}=k_c/\sqrt{3}; \quad
\tilde{\tau}= \Big\{ \frac{3^{3/2}\hbar}{ 4m(g^3\delta)^{1/2}}
 \Big\}^{1/2}\Big(\frac{\rho_1+\rho_2}{\rho_1-\rho_2}\Big)^{1/4}.
\label{taufast}
\eeo

\subsection{Future Evolution of the RT Instability}

We expect the instability to progress into the nonlinear regime
through the development of spikes and bubbles \cite{baym,sharp}.
Spikes longer than their diameters are unstable to varicose
perturbations leading to their breakup into droplets. Spike
breakup can also occur due to the Kelvin-Helmholtz Instability at
the edges of the spikes \cite{sharp} in both 2D and 3D cases.
Spike breakup leads to droplet formation in both cases instead of
a mixing zone when surface tension is significant, since droplets
minimize surface energy. The diameter of the droplets is then half
the wavelength of the fastest growing mode, $\tilde{\lambda}/2$.

For low non-zero temperatures, the motion of the condensate and
the excitations are decoupled. Since the excitations are dilute
and of long wavelength, the exciton fluid continues to be RT
unstable with the droplet size determined by the effective tension
of the condensate and the applied stress or electric field
gradient. For weak interactions and low temperatures, $ \rho_0
U_0<< T << T_c$, the condensate moves inviscidly in a mean field
due to the excitations in the Hartree-Fock approximation. If the
external potential is modified to include the mean field due to
excitations as a correction, the discussion in Sec. III can be
used to show that the BEC is RT unstable with the applied force
and the effective tension again determining the droplet size. At
$T>T_c$, however, when the fluid is normal and has viscosity, the
character of the flow changes. Though the fluid can still be RT
unstable, the droplet size is now determined by the viscosity and
applied force. Therefore, one can observe in experiments the
variation in droplet size as T is increased beyond $T_c$.

\section{Application to Special Cases}

Below we investigate the effects of the Rayleigh-Taylor
instability in two systems of special interest.

\subsection{Atomic BECs}

Consider a typical experiment in which we have a BEC of alkali
atoms in a three dimensional magnetic trap. When the trap
potential is switched off, the condensate is in a non-equilibrium
state. Repulsive interactions make the condensate expand slowly.
After a a time of flight of several nanoseconds, a snapshot of the
freely expanding BEC is usually taken. We propose one modification
of the experiment which can lead to the formation of droplets of
BEC at the surface of the condensate while the whole gas expands.
We suggest that after turning off the magnetic trap, we turn on an
``inverted magnetic trap" with a potential profile, $V(\vec{r})
\sim -1/2 ~ w^2 r^2$, resulting in a radial outward force acting
on the expanding condensate. This outward force will result in a
Rayleigh-Taylor instability at the surface of the expanding
condensate.

It is simple to generalize to three dimensions the results of Sec.
III. We continue to choose y as the linear axis perpendicular to
the surface. In 3D, the Eqs.\,(\ref{eee}),
(\ref{inst})-(\ref{deltaA}) and the later expressions for the
growth time and wavelength, Eqs. (\ref{tau}) and (\ref{taufast})
hold exactly with $k$ now being given by $k^2=k_x^2+k_y^2+k_z^2$,
and every perturbed quantity, $Q$, growing in time as,
\begin{equation}
 \delta Q(x,y,t) = Q'(y) e^{ik_x x + i k_y y + i k_z z+ t/\tau}.
\end{equation}

For $^{87}$Rb atoms, a magnetic trap of frequency, $24-240$ Hz can
exert a restoring force equivalent to an acceleration of $g \sim
4.55-455$ cm/s$^2$, at a distance of 2 $\mu$m away from the trap
center. A BEC contained in a trap of frequency, $24$ Hz will have
a radius $\sim 2 \mu$m. After the BEC is released from this trap,
an inhomogeneous magnetic field can be turned on so that the force
exerted points outwards from the center instead of inwards. The
BEC, which is now in a nonequilibrium state will then begin to
expand. If the force of expansion is of the appropriate magnitude,
there will be an RT instability  driven by the force of expansion
which will lead to the formation of RT droplets. A strong outward
force corresponding to an acceleration, $g \sim 5 \times 10^5$
cm/s$^2$, leads to a fastest growing wavelength of 0.1 $\mu$m (Eq.
\ref{taufast}), if we assume that the surface width, $\delta$, is
almost the same in magnitude as the wavelength. This leads to
droplets of size, $>0.05$ $ \mu$m, which can be observed. The
growth time for such a droplet is given by Eq. \ref{taufast} to be
$\tilde{\tau} \sim 3$ $ \mu$s. Weaker forces leads to larger
wavelengths for the instability. Since for an observable
instability, we need the wavelength, $\tilde{\lambda} < \pi D$,
where $D$ is the diameter of the cloud, we find that the maximum
value of acceleration at the surface is, $g_{max}=0.8$ cm/s$^2$,
which leads to a minimum growth time for observable instability,
$\tilde{\tau}_{min} \sim 0.02$ sec for $\tilde{\lambda}_{max}
\simeq D \sim 4$ $ \mu$m (Eq. \ref{taufast}).

A suitably designed experiment can therefore be used to create
Rayleigh-Taylor droplets in an atomic BEC. In an actual
experiment, the condensate will be allowed to expand to several
times its initial size before measurements are made. Therefore,
the droplets seen at measurement will be scaled up.

\subsection{Excitonic BEC in Layered Semiconductors?}

The possibility exists that an indirect exciton cloud in a layered
semiconductor if cooled below Tc may form two dimensional BECs
trapped in impurity driven in-plane shallow traps
\cite{butov2,butov1}. This type of BEC has not yet been
conclusively observed. If, however, such a BEC can be created in
the future, it may be possible to overturn the trap by applying an
outward radial force through an inhomogeneous stress. This would
lead to a two-dimensional BEC in a non-equilibrium state, which is
driven to expand by the radial force resulting in an RT
instability. We note here that inhomogeneous stress has previously
been used to create a harmonic trap in a semiconductor bilayer
\cite{negoita}. It is reasonable to assume that it may be possible
to use the same means to create an outward, instead of inward,
radial force field.

 For indirect excitons, $m=0.1 m_e$, $\rho \sim 10^{16}$
cm$^{-3}$ in the wells, and $m s^2 \sim \rho U_0 \simeq 0.1-1$ meV
\cite{butov2}. If $g=10^{15}$ cm s$^{-2}$ as achieved in Ref.
\cite{negoita} through inhomogeneous stress and electric fields,
and making the reasonable assumption that $\delta \sim
\tilde{\lambda}$, under which conditions the RT instability still
exists, we find $\tilde{\lambda} \sim 5.4$ $\mu$m and
$\tilde{\tau} \sim 0.4$ ns. Moreover, it can be verified that $s
\gtrsim \bar{\lambda}/\bar{\tau}$. Therefore, we would expect BEC
droplets, 5-10 microns in diameter, but as the exciton cloud
starts to expand radially, the size of the droplets will scale
accordingly.

\section{Conclusions}

While different statistical distributions are used to correlate
data on the drop size in normal fluids, such as the
Nukiyama-Tanasawa law \cite{sharp,nukiyama}, which is arrived at
through assumptions on droplet formation, a theory for the
distribution in the droplet size in a quantum fluid such as a BEC
needs to be developed. We also need new experiments to study BEC
droplets to further our understanding of the quantum mechanics.

\section{Acknowledgements}

The author wishes to thank G. Baym and J.P. Wolfe for helpful
discussions on the Rayleigh-Taylor instability in excitons.

\end{document}